\documentclass[10pt,conference,letterpaper]{IEEEtran}
\IEEEoverridecommandlockouts
\usepackage{cite}
\usepackage{amsmath,amssymb,amsfonts}
\usepackage{algorithmic}

\usepackage{algorithm2e}

\RestyleAlgo{ruled}
\usepackage{graphicx}
\usepackage{textcomp}
\usepackage{xcolor}
\usepackage{caption}
\usepackage{subcaption}
\usepackage{enumitem}
\PassOptionsToPackage{hyphenat}{url}
\usepackage[none]{hyphenat}
\def\BibTeX{{\rm B\kern-.05em{\sc i\kern-.025em b}\kern-.08em
    T\kern-.1667em\lower.7ex\hbox{E}\kern-.125emX}}
\begin{document}

\newcommand{\ek}[1]{{\color{blue}[\textbf{\sc ek}: \textit{#1}]}}
\newcommand{\hc}[1]{{\color{red}[\textbf{\sc hc}: \textit{#1}]}}
\makeatletter 
\newcommand{\linebreakand}{%
  \end{@IEEEauthorhalign}
  \hfill\mbox{}\par
  \mbox{}\hfill\begin{@IEEEauthorhalign}
}
\makeatother 
\title{A Robust Power Model Training Framework for Cloud Native Runtime Energy Metric Exporter}
\author{
\IEEEauthorblockN{Sunyanan Choochotkaew}
\IEEEauthorblockA{\textit{{IBM Research}}\\
Tokyo, Japan\\
Sunyanan.Choochotkaew1@ibm.com}
\and
\IEEEauthorblockN{Chen Wang}
\IEEEauthorblockA{\textit{IBM Research} \\
Yorktown Heights, U.S.A.\\
chen.wang1@ibm.com}
\and
\IEEEauthorblockN{Huamin Chen}
\IEEEauthorblockA{\textit{Red Hat Inc.}\\
Boston, U.S.A.\\
hchen@redhat.com}
\linebreakand
\and
\IEEEauthorblockN{Tatsuhiro Chiba}
\IEEEauthorblockA{\textit{IBM Research}\\
Tokyo, Japan \\
chiba@jp.ibm.com}
\and
\IEEEauthorblockN{Marcelo Amaral}
\IEEEauthorblockA{\textit{IBM Research} \\
Tokyo, Japan \\
marcelo.amaral1@ibm.com}
\and
\IEEEauthorblockN{Eun Kyung Lee}
\IEEEauthorblockA{\textit{IBM Research} \\
Yorktown Heights, U.S.A. \\
eunkyung.lee@us.ibm.com}
\and
\IEEEauthorblockN{Tamar Eilam}
\IEEEauthorblockA{\textit{IBM Research} \\
Yorktown Heights, U.S.A. \\
eilamt@us.ibm.com}
}

\maketitle

\begin{abstract}
Estimating power consumption in modern Cloud environments is essential for carbon quantification toward green computing. Specifically, it is important to properly account for the power consumed by each of the running applications, which are packaged as containers. This paper examines multiple challenges associated with this goal. The first challenge is that multiple customers are sharing the same hardware platform (multi-tenancy), where information on the physical servers is mostly obscured. The second challenge is the overhead in power consumption that the Cloud platform control plane induces. This paper addresses these challenges and introduces a novel pipeline framework for power model training. This allows versatile power consumption approximation of individual containers on the basis of available performance counters and other metrics. The proposed model utilizes machine learning techniques to predict the power consumed by the control plane and associated processes, and uses it for isolating the power consumed by the user containers, from the server power consumption. To determine how well the prediction results in an isolation, we introduce a metric termed \emph{isolation goodness}. Applying the proposed power model does not require online power measurements, nor does it need information on the physical servers, configuration, or information on other tenants sharing the same machine. The results of cross-workload, cross-platform experiments demonstrated the higher accuracy of the proposed model when predicting power consumption of unseen containers on unknown platforms, including on virtual machines. 
\end{abstract}

\begin{IEEEkeywords}
Cloud, Green computing, Power model, Containers, Machine learning
\end{IEEEkeywords}
\section{Introduction}
\label{sec:intro}
Towards green computing in modern Cloud environments, quantifying the energy consumption by utilizing a fine-grained container power model can help with both transparency and awareness. In addition, such a quantification is an important building block for developers to enhance their codes, and for administrators to enable intelligent resource management systems for optimizing the energy consumption in a {\em container orchestration platform} (such as Kubernetes), similarly to what have been done in non-container systems \cite{wl_aware, energy_aware, energy_eff, treehouse}.

{\em Server power consumption} is the power consumed by a physical server machine to run logical processes. The current finest granularity which is physically measurable is on a system-on-chip (SoC) domain with onboard power meter capability integration \cite{sandy}. There have been multiple attempts to logically decompose the measured power into units that correspond to individual processes that are running simultaneously by leveraging machine learning techniques \cite{taxonomy}. 

One of the challenges in such modeling attempts is that the power consumption of identical machines may differ, even if they are running identical programs with an identical load. These differences stem from physical factors such as CPU architectures, and ambient temperature, and from logical factors such as CPU frequency, and operating systems \cite{physicalvariation}. Traditionally, the power consumption of background processes at the idling state before running the workloads is nearly static. Thus, a common approach to isolate the workload power consumption is to profile the power at the idling state and deduct it from the measured power \cite{vm-model}.

However, there are two serious impediments associated with the profiling approaches to train a power model for containers running on multiple-tenant container orchestration systems managed by a Cloud environment (e.g., Amazon EKS\cite{eks}, Google GKE \cite{gke}, or IBM Cloud IKS\cite{iks}). Firstly, a server usually has multiple virtual machines (VMs), in which workload containers are located, running at the same time. The power measurements are different when changing the server CPU frequency governor or the number of co-located VMs. Lacking access to these data (i.e., CPU frequency and co-locating VMs) due to the multiple-tenant resource sharing nature hampers the ability of the profiling approach to apply the matching profile.
Secondly, the container orchestration systems comprise multiple control plane processes for handling the container stack (e.g., networks and storage) along with the entire life cycle management of containers including deployment and placement. We can expect that more background processes will become active upon any life cycle event, e.g., the starting of a container. As a result, the power consumed by background processes is not static and can be noisy (see, Fig.~\ref{fig:noisy}), which poses complications for container power modeling, and in particular for the use of static profiling. 
Unlike modeling operating system processes \cite{runtime_os}, it is a huge and seemingly endless task to model an extensible set of control plane processes.

\begin{figure}
\centering
\begin{subfigure}[b]{0.22\textwidth}
\centering
    \includegraphics[width=1\linewidth]{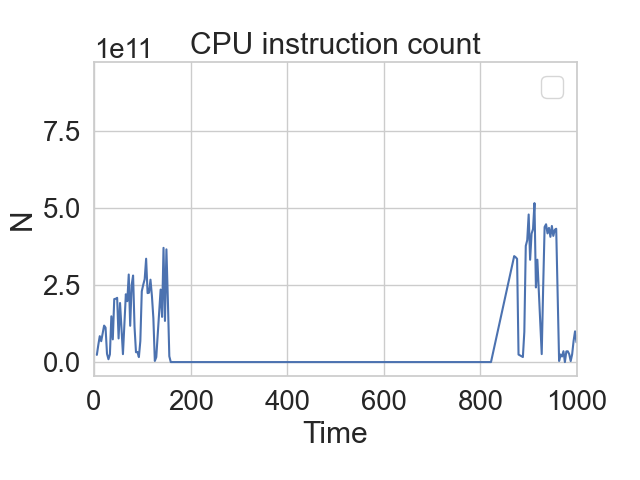}
    \subcaption{Collected workload usage}
    \label{fig:noisyusage}
\end{subfigure}
\hfill
\begin{subfigure}[b]{0.22\textwidth}
\centering
    \includegraphics[width=1\linewidth]{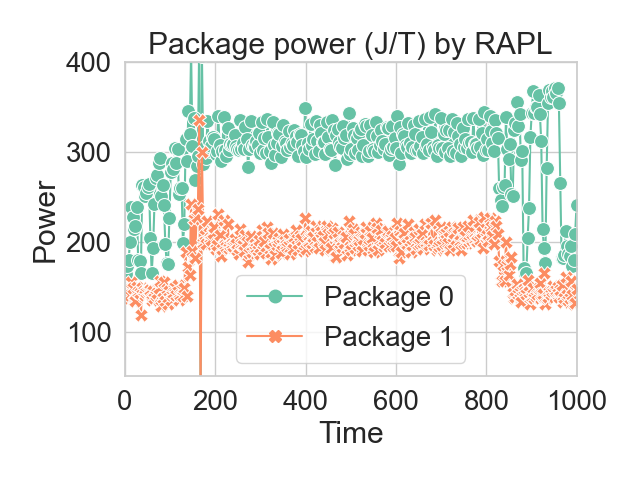}
    \subcaption{Measured power}
    \label{fig:noisypower}
\end{subfigure}
\caption{(a) Workload usage is not always correlated to (b) power consumption due to noisy background processes.}
\label{fig:noisy}
\end{figure}

\begin{figure}[t!]
    \centering
    \includegraphics[width=0.35\textwidth]{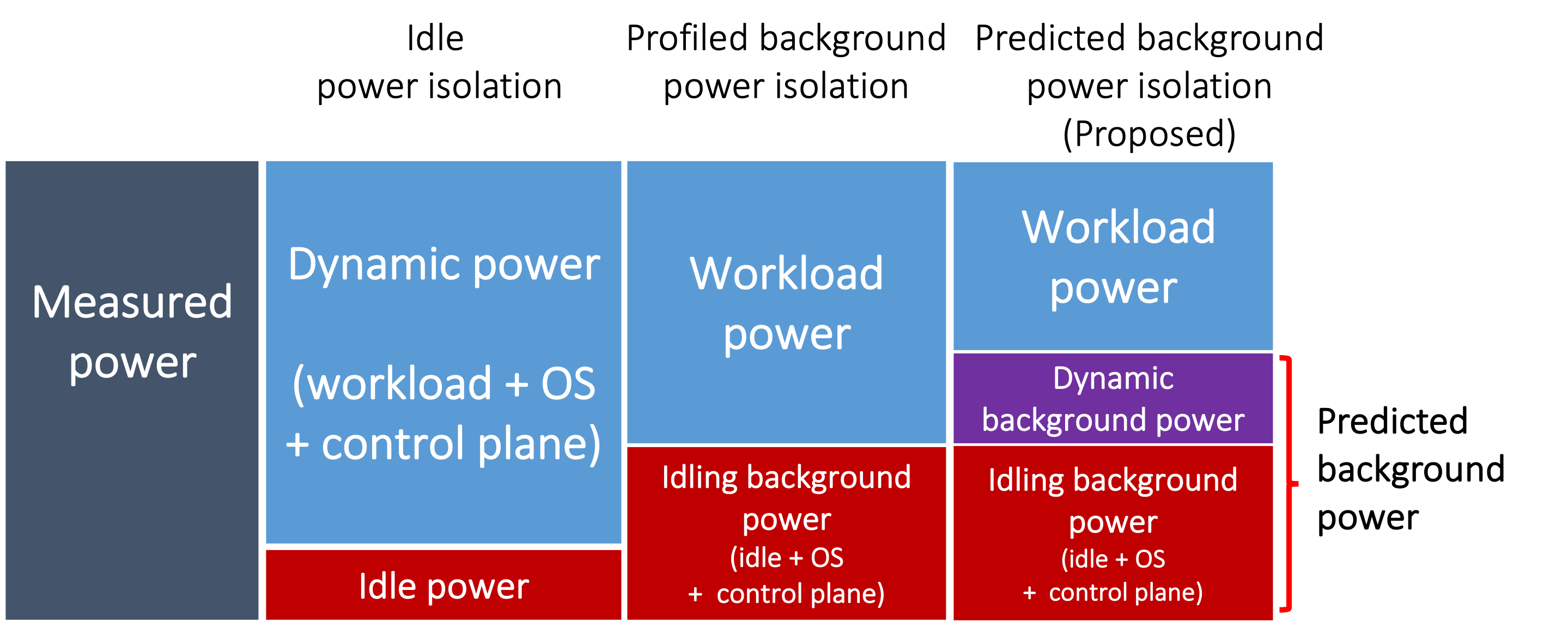}
    \caption{Dynamic power isolation for model training.}
    \label{fig:concept}
\end{figure}

In response to these challenges, we introduced a versatile pipeline framework of power model training integrated to the model server project of Kepler, a fine-grained energy-related metric exporter on Cloud \cite{kepler}. Kepler collects resource usage and energy metrics from multiple sources, termed metric producers, and exports them to the metric server for both server level and container level. The proposed pipeline allows training a power model for both levels with respective to available usage metrics exported by the exporter. Specifically, we highlight a power isolation module, a key process to divide the measured server power consumption into a power consumption stems from the resource usage of each container. Once trained, the model does not need any platform data, nor does it require power measurements. 

Typical power isolation methods separate the measured power into \textit{(idling) background power} and \textit{workload power} (Fig.~\ref{fig:concept}), where the existing definition of background power is the power measured before running the workload (i.e., at idling state), and the workload power is the power difference after becoming stressed by the workload. In contrast to a non-orchestrated platform, the resource usage of background processes can be varied when running workload containers. As an extension of our previous short paper \cite{model_server_mascot}, we redefine this variation as \textit{dynamic background power} and propose a new isolation method considering the dynamic background power to improve an accuracy of power consumption estimation, regardless of the operating platform conditions (e.g., platform-specific background processes). Furthermore, the proposed isolation method does not require extensive profiling, thus reducing the amount of work and the complexity. This paper elaborates our contributions as follows.
\begin{itemize}
    \item \textbf{Versatile pipeline framework for power model training}: We introduce a pipeline framework which enables the power of crowd-sourcing to build a power model from any emerging containerized benchmarks by using any learning approaches with resource usage metrics that are available on their platforms.

    \item \textbf{Novel approach of power isolation}: In addition to idling background power, the proposed isolation method uses a machine learning technique to estimate the dynamic portion of background power (i.e., dynamic background power) and remove it from the power model training. 

    \item \textbf{Intuitive goodness definition of power isolation}: We define a metric to evaluate how well the isolation method extracts a workload power portion based on a correlation value with the resource usage.
 
    \item \textbf{Cross-workload, cross-platform validation}: We present the results of our cross-validation experiments consisting of multiple workloads with different operating requirements, and multiple platforms with different CPU frequency governors \cite{cpugovernor}, and different virtualization.
\end{itemize}

Section~\ref{sec:problem} of this paper clarifies the problem definition. Section~\ref{sec:related} presents related works. In Section~\ref{sec:model}, we explain our modeling approach, and in Section~\ref{sec:experiment}, we present the evaluation and validation results. We conclude in Section~\ref{sec:conclude} with a brief summary and mention of future work.
\section{Problem Definition}
\label{sec:problem}
To train a power model, processes are separately considered as either \textit{workload} or \textit{background}. \textit{Workload} refers to a benchmark process that the model builder runs to stress the machine for a high power variation in the training data acquisition phase. \textit{Background} refers to the other processes including operating system (OS) and control plane. The high correlation between the power and resource usage of the workload is easily observed, but even so, there are multiple factors causing a variation on the absolute power number which cannot be profiled in advance, especially background processes. The dynamicity of background processes potentially raises the following power modeling and carbon accounting concerns and challenges.

\begin{enumerate}[label=P-\Roman*]
    \item A model trained on one platform environment is degraded when applied to another platform environment. 
    \item A model trained by one kind of container workload is degraded in accuracy when applied to a different kind of container workload.
    \item Co-located containers that did not exist in the training phase degrade the model accuracy.
    \item Even though some portion of the increased power consumption after running the workload is consumed by background processes of the platform, this portion has never been identified and reported to the providers.
\end{enumerate}
\section{Related Works}
\label{sec:related}
The significant amount of energy consumed by data centers has been a topic of research interest for more than a decade, with a variety of studies investigating fine-grained energy and power modeling in virtualization architectures such as virtual machines (VM) and containers \cite{taxonomy}.

Server power consumption can be measured from power meter instrumentation \cite{powermon}, from dedicated acquisition systems \cite{ipmi}, and from a software power meter \cite{rapl}. The most extensively utilized power meter is the \emph{running average power limit (RAPL)}\cite{rapl} software power meter. For modeling the container power, there are two common approaches. The first is to assume that the power measurement is obtainable on estimation.  The measured power is distributed to the container powers using the ratio of its resource utilization over the total utilization \cite{deepmon}. The second approach is to assume there is no power meter on estimation, which is considered as \emph{non-RAPL}. In most Cloud environments, the power meter is not accessible, so we focus on the second approach in this paper.

\subsection{Machine learning approach and features}
\label{subsec:learning}
Machine learning and other statistical methodologies have high potential to infer the container-level power from its virtually countable resource usage events for the non-RAPL assumption. A high correlation between the resource usage obtained from  hardware counters, and the measured power has been reported in several studies \cite{power_hpc, pmc_energy, runtime_os, lightweight, intelxscale, selfwatts, tenml}.

\begin{figure}
\begin{subfigure}[b]{0.24\textwidth}
\centering
    \includegraphics[width=1\linewidth]{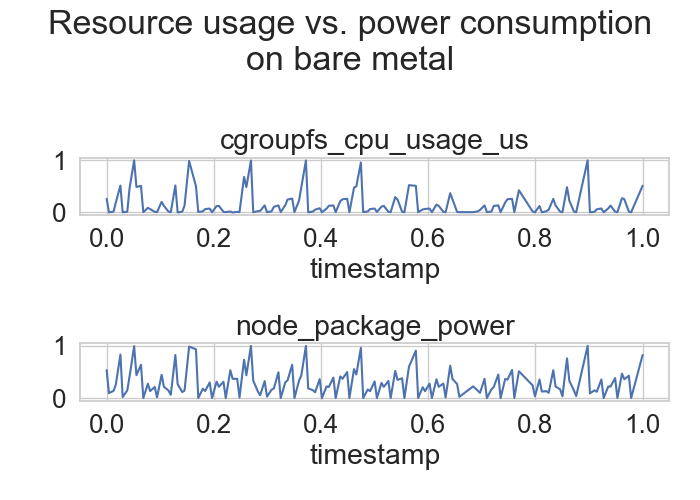}
    \subcaption{on bare metal}
    \label{fig:snapshot_on_bm}
\end{subfigure}
\hfill
\begin{subfigure}[b]{0.24\textwidth}
\centering
    \includegraphics[width=1\linewidth]{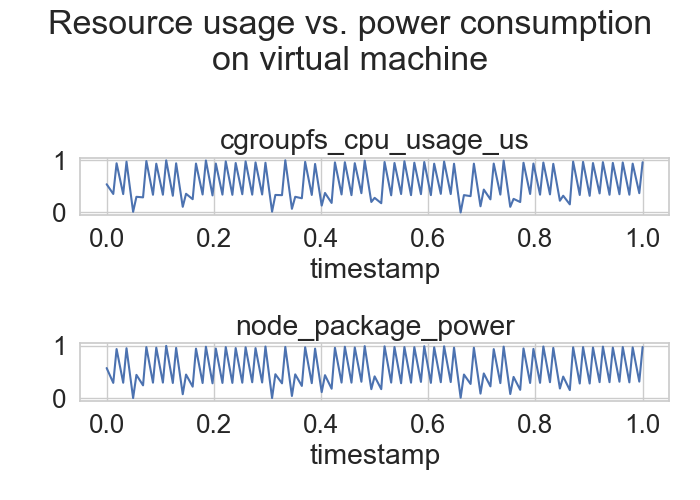}
    \subcaption{on virtual machine}
    \label{fig:snapshot_on_vm}
\end{subfigure}
\caption{Snapshot of normalized Kepler metrics showing high correlation between resource usage and power consumption when running Coremark benchmark.}
\label{fig:snapshot}
\end{figure}

\begin{figure}[!t]
    \centering
    \includegraphics[width=0.24\textwidth]{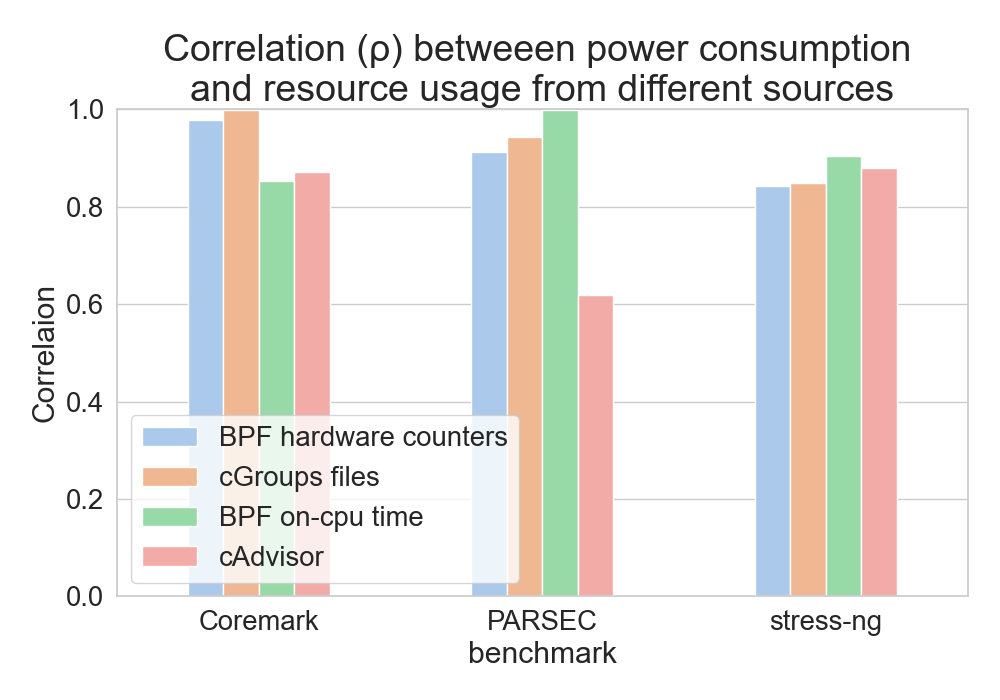}
    \caption{Correlation between resource usage from different metric producers and RAPL power for each benchmark.}
    \label{fig:correlation}
\end{figure}

Likewise, we observed a high similarity between resource usage and power consumption from the data snapshot not only when running a workload on the bare metal but also when running on the virtual machine as shown in Fig.~\ref{fig:snapshot}. 
Furthermore, according to our experiments, not limited to hardware counters, the other resource usage metrics exported by Kepler such as CPU time from cGroups \cite{namespaces}, cAdvisor \cite{cadvisor}, and BPF probe \cite{ebpf_perf} also have a high correlation to the measured power on our benchmarks as shown in Fig.~\ref{fig:correlation}. 

For learning approaches, the simplest approach is to utilize linear regression \cite{lr}, as demonstrated in \cite{energy_in_docker, vm-model, power_hpc}. Tadesse et al. used polynomial regression in their container power model \cite{energy_in_docker}, which was also utilized in \cite{process-level} in the context of general processes. SmartWatts\cite{smartwatts} utilized ridge regression \cite{ridge}. There have also been power modeling studies using the other regression-based learning approaches such as Gaussian boosting (GB), support vector machine (SVM), and k-nearest neighbors (kNN) in \cite{tenml, vm-model}, as well as a neural network such as multi-layer perceptron (MLP) in \cite{cloudml}. However, as yet there is no consensus on the best learning approach for power model training.

\subsection{Power isolation and model labeling}
\label{subsec:isolate}
Multiple factors can introduce complications to the power model training in non-RAPL approaches. Some are physical factors such as CPU architecture, manufacturing, ambient temperature \cite{physicalvariation}, while others are logical factors such as CPU frequency governor settings, operating systems, and control plane processes. There are multiple ways to extract only dynamic changes from the measured powers when stressing the system with a workload.
Conventionally, the measured power is decomposed into \textit{idle power} and \textit{dynamic power}. \textit{Idle power} is a static energy consumption due to current leakage at the state where no process is running. \textit{Dynamic power} is the remaining part of the measured power after removing the idle power. Correspondingly, the container power model is trained by all running processes \cite{process-level}.
Some modeling approaches collect the power measurement before running the workload process, defined as \textit{idling background power}. Then, the container power model is trained by the power difference between the measured power when running the workload and the idling background power. The broad differentiation of the proposed power modeling is illustrated in Fig.~\ref{fig:proposed_concept}. 

As for profiling-based approaches, Containergy \cite{containergy} is a framework to generate containerized workload profiles of usage metrics from hardware counters over energy for each controlled CPU frequency setting. However, these profiles do not consider applications to unseen containers.
cWatts++ \cite{lightweight} has eventModel for non-RAPL container power modeling, where the background power is estimated with a quadratic function of CPU frequency using profiled coefficients. Considering unseen containers, cWatts++ trains the model with various workloads from the PARSEC benchmark suite \cite{parsec} to minimize the effect of workload bias \textit{P-II} stated in Section~\ref{sec:problem}. Similarly, SmartWatts \cite{smartwatts} utilizes a machine learning technique to build a model for a given frequency. SmartWatts assumes an online calibration mechanism with consideration of both platform and workload bias \textit{P-I and P-II}.

The proposed container power model not only does away with the power meter and profiling requirements but also excludes the training biases \textit{P-I}, \textit{P-II}, and \textit{P-III} and identifies the dynamic background power portion to be considered on the platform provider (\textit{P-IV}). 

 \begin{figure}[t!]
    \centering
    \includegraphics[width=0.4\textwidth]{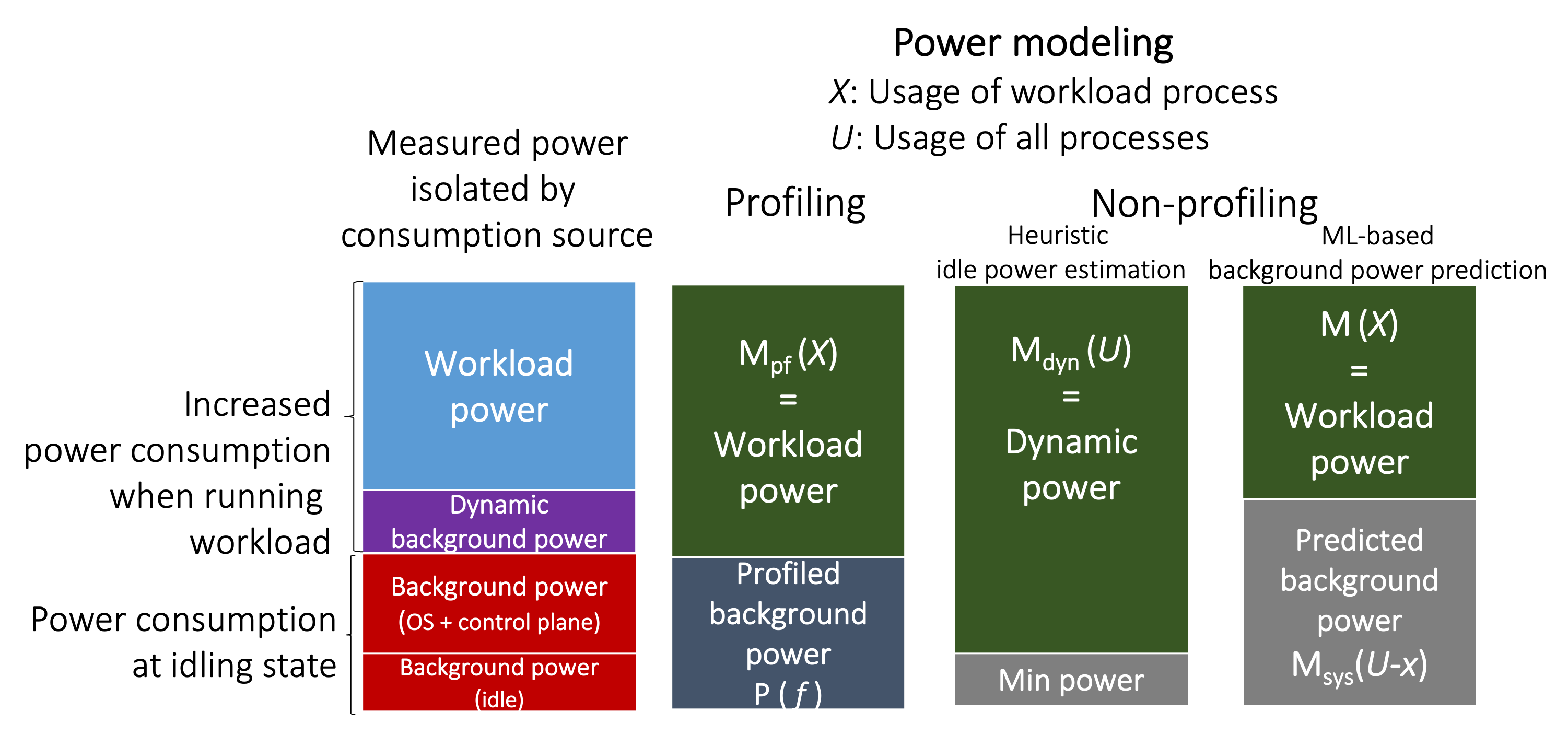}
    \caption{Non-RAPL power modeling.}
    \label{fig:proposed_concept}
\end{figure}
\section{Power model training pipeline framework}
\label{sec:model}

\begin{figure}[t!]
    \centering
    \includegraphics[width=0.45\textwidth]{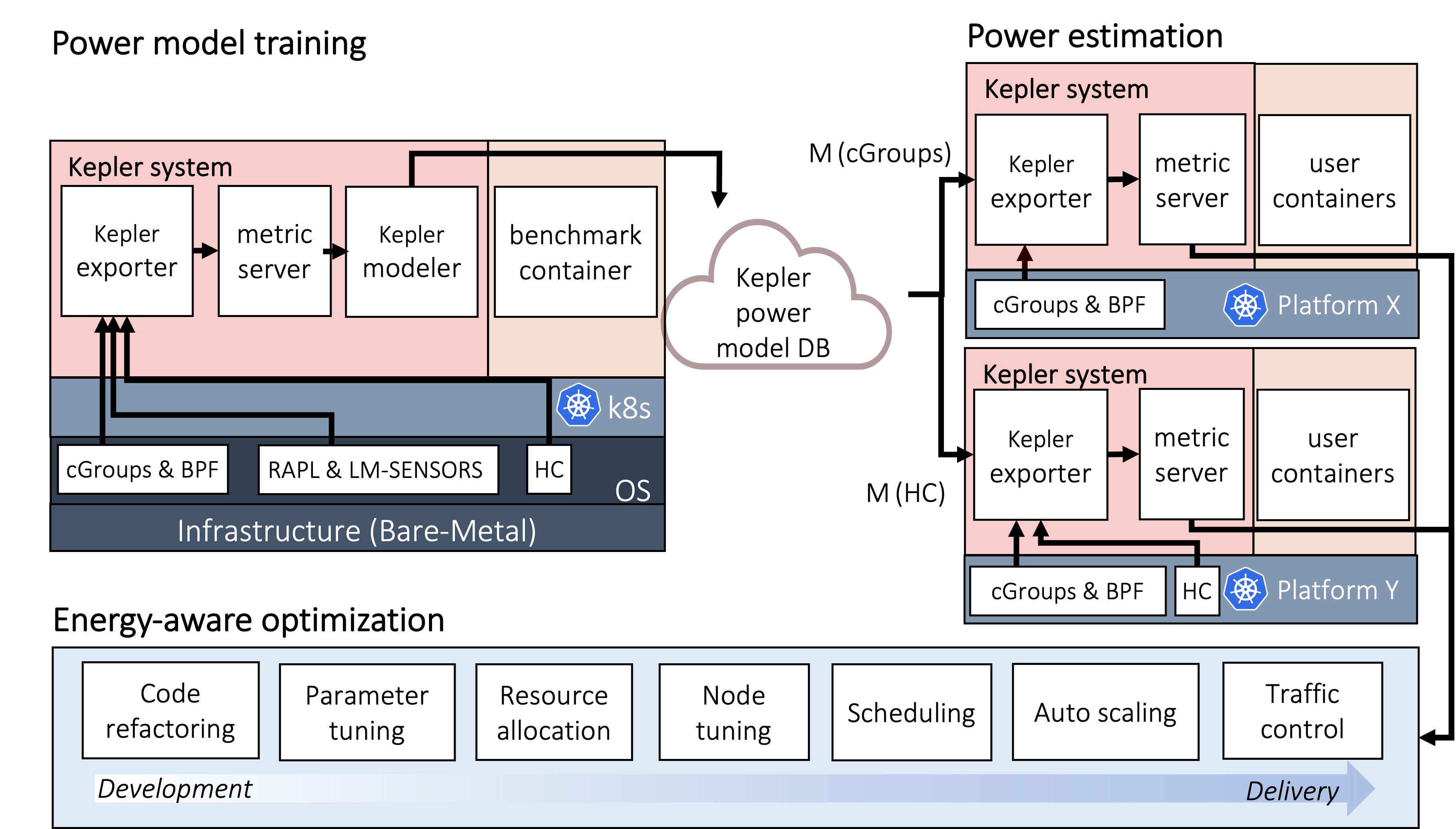}
    \caption{Integrated modeler module on Kepler toward a sustainable Cloud.}
    \label{fig:kepler_arch}
\end{figure}

\begin{figure}[t!]
    \centering
    \includegraphics[width=0.45\textwidth]{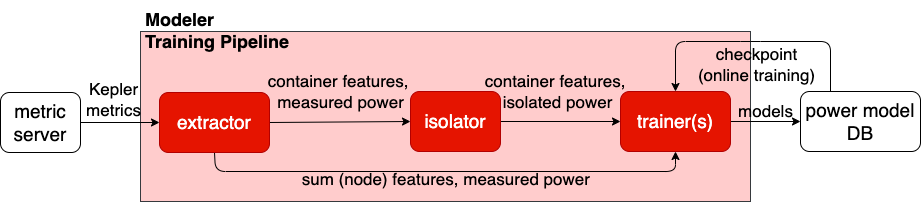}
    \caption{Training pipeline of Kepler modeler.}
    \label{fig:kepler_pipeline}
\end{figure}

\begin{figure}[t!]
    \centering
    \includegraphics[width=0.35\textwidth]{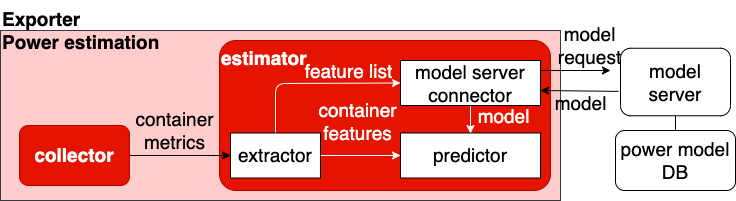}
    \caption{Power estimation process.}
    \label{fig:operate_flow}
\end{figure}

The proposed pipeline framework for power model training is integrated to the model server component of \emph{Kepler}, an open source project to export energy-related metrics in a Cloud-native manner \cite{kepler}. This integration allows building a container power models via crowd-sourcing with any emerging benchmarks on any training platforms running Kepler and using these models to predict the power consumption of unseen containers in unknown platforms on a basis of available performance counter and other metrics, as illustrated in Fig.~\ref{fig:kepler_arch}. We assume here that power model contributors run workload containers on their platforms and share power-related metrics (including the measured power) to a metric server via the exporter. With different sets of features grouped by the metric producer, the modeler performs a training pipeline process and saves the model is the model database. Then, Kepler on the platforms that have no access to an online power measurement applies the container power model depending on the availability of the metric producers to predict container power consumption. The quantified energies, i.e., total predicted powers, of user containers can be employed in various ways for energy-aware optimization throughout the software development and delivery lifecycle.

The training pipeline is composed of three modules: extractor, isolator, and trainer as shown in Fig.~\ref{fig:kepler_pipeline}. From Kepler, we use energy metrics at the node level, and resource usage metrics at the container level as inputs. We consider a node as a system. \emph{Extractor} module pre-processes the Kepler metrics by transforming accumulated energy to watt unit, transforming accumulated resource usage metric to per-second value, grouping the resource usage metric by metric producer, and cleaning the data. The output are the container features for each metric producer group labeled with system-level measured power at a specific second. The aggregated usage value of all containers can be used to train a system-level power model. Meanwhile, the per-container usage value is further submitted to isolator module. \emph{Isolator} calculates isolated power which is a remaining power after excluding platform-specific power (e.g., idle power, idling background power, and dynamic background power) from the measured power and submits to trainer module. \emph{Trainer} module can have more than one machine learning approaches to build multiple candidate power models since there are several factors that result in one approach outperforming the other such as the amount of data. The candidate models are then saved to power model database. The pipeline also allows an online training by loading a checkpoint from the database. As a result, Kepler exporter can select the currently best-performing power model via the model server as shown in Fig.~\ref{fig:operate_flow}.

\begin{figure}[t!]
    \centering
    \includegraphics[width=0.4\textwidth]{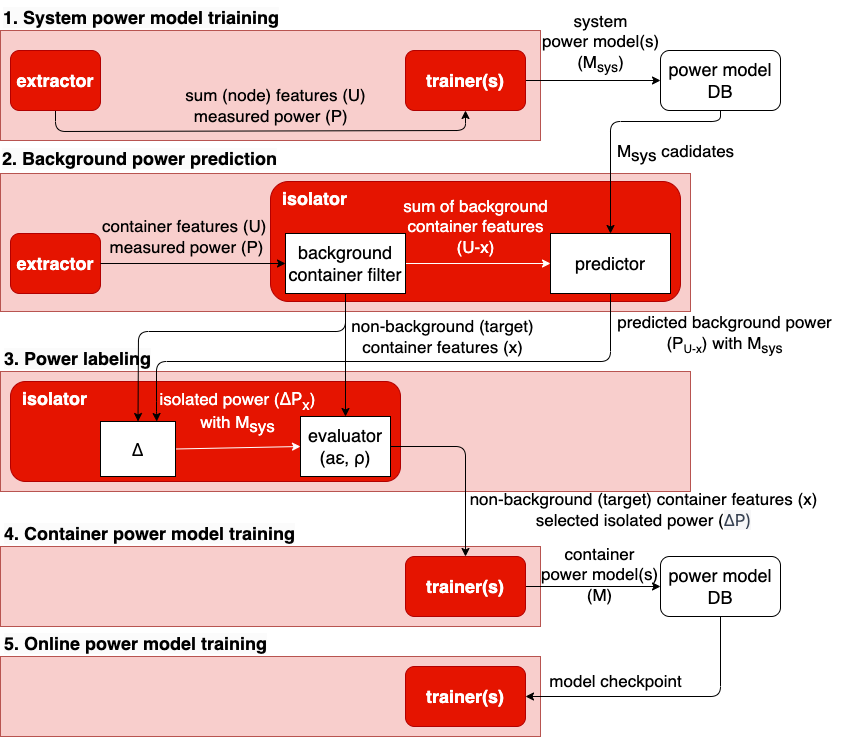}
    \caption{Training pipeline with the proposed isolator.}
    \label{fig:train_flow}
\end{figure}

\subsection{Training pipeline with the proposed power isolation}
The proposed training pipeline comprises five steps, as shown in Fig.~\ref{fig:train_flow}. Step~$1$ trains the system power model candidates, denoted as $M_{sys}$. Step~$2$ predicts the background power, which is done by running each candidate model on the background containers. Step~$3$ evaluates and selects the best isolated power based on the candidate model accuracy ($a\epsilon$) and  the newly defined isolation goodness ($\rho$). Step~$4$ trains the container power model. Lastly, Step~$5$ performs online training upon a new batch of collected data. 

\subsubsection{Step 1: System power model training}
The \emph{system power model} ($M_{sys}$) candidates are trained by using the aggregated resource usage from all containers as features and the measured power as labels. Given the time series of a server power consumption measured by a power meter ($P:\{p_i,...,p_n\}$) and those of an aggregated resource usage ($U:\{u_i,...,u_n\}$), a system power model is fit and evaluated with the mean absolute error as below.

\begin{equation}
\label{eq:abs_err}
\begin{split}
    P = &M_{sys}(U) + a\varepsilon \\
    M_{sys} \text{ error } (a\varepsilon) =& \frac{\sum_{i=1}^n|p_i - M_{sys}(u_i)|}{n}\\
\end{split}
\end{equation}

\subsubsection{Step $2$: Background power prediction}
At this step, the containers are separated into two groups: background containers and non-background (target) containers. The background container filtering can be done in several ways. The most trivial algorithm is to define a list of background containers. With the system models ($M_{sys}$) from Step~$1$, the background power ($P_{U-x}$) is predicted by the aggregated resource usage ($U$) deducted by the resource usage of target containers ($x$). 
\begin{equation}
P_{U-x}=M_{sys}(U-x)
\end{equation}

\subsubsection{Step~$3$: Power labeling}
The difference between the measured power and the predicted background power is the workload power, denoted as $\Delta P_{x}$. 
\begin{equation}
\label{eq:px}
\Delta P_{x}=P-P_{U-x}
\end{equation}

The best workload power labels from each contributor platform (e.g., $\Delta P$) are determined by two metrics: (i) model accuracy in Equation~(\ref{eq:abs_err}) and (ii) isolation goodness which is newly defined as follow. 

Given $F:\{f\}$ as a set of features of the considering metric producer group, $x_{f}(t)$ as the $f$ value of the container $x$ at time $t$, and $\Delta P_x(t)$ as the isolated workload power at time $t$, we define an isolation goodness ($\rho$) with the Pearson correlation coefficient, as below. 

\begin{equation}
\label{eq:corr}
\begin{split}
\text{Isolation goodness }(\rho) = &max_{f\in F}\ corr(x_f(t), \Delta P_x(t));\\
corr(I:\{i\}, J:\{j\}) =& \frac{\sum(i-\overline{i})(j-\overline{j})}{\sqrt{\sum(i-\overline{i})^2\sum(j-\overline{j})^2}}
\end{split}
\end{equation}

The power model labeling is performed using Algorithm \ref{algo:select}. For each system model candidate $m$ in $M_{sys}$, we estimate the workload power ($\Delta P_x$) from Equation~(\ref{eq:px}) and compute the isolation goodness ($\rho$) from Equation~(\ref{eq:corr}). Given an acceptable threshold of the isolation goodness $\rho_{th}$, $m$ is considered as a better candidate if it satisfies either of the following conditions: (i) there is no other candidate, (ii) $m$ has lower error ($a\varepsilon$) and $\rho$ is acceptable, (iii) the comparing candidate $M_{best}$ does not satisfy $\rho_{th}$ and $m$ has a lower error with a higher or equal isolation goodness. The system power model error generally varies due to the platform complexity, such as the number and dynamicity of control plane containers or co-locating virtual machines. System power models with a high error usually result in low isolation goodness. For preliminary investigation, we empirically set an acceptable threshold to 0.7. Power labels from the best candidate are used for the next step.

\subsubsection{Step $4$: Container power model training}
For each learning approach in trainer module, the \emph{container power model} ($M$) is trained by fitting an aggregated resource usage from target containers ($x: {x_i}$) as features and the isolated power ($\Delta P: {\Delta p_i}$) from Step~$3$ as labels and is evaluated with the mean absolute error as below, where $n$ is the number of data points in the collected time series.

\begin{equation}
\label{eq:dyn_err}
\begin{split}
    \Delta P = &M(x) + d\varepsilon \\
    M \text{ error } (d\varepsilon) =& \frac{\sum_{i=1}^n|\Delta p_i - M(x_i)|}{n}\\
\end{split}
\end{equation}

If the measured power ($P$) and idling background power ($P_{profile}$) are available, the dynamic background power ($\Delta P_{bg}$) can be approximated by this container power model ($M$) as below.

\begin{equation}
\label{eq:pxop}
\Delta P_{bg} \approx P - P_{profile} - M(x)
\end{equation}

\subsubsection{Step $5$: Online power model training}
When a new batch of data is fed to the pipeline, Step~$1$ to Step~$3$ are repeated. Then, at Step~$4$, a checkpoint from the previous training is loaded for incremental training. 

\begin{algorithm}[t!]
  \SetAlgoLined
  \KwData{P, U, x, $M_{sys}$, $\rho_{th}$}
  \KwResult{$\Delta P$}
  $M_{best}\leftarrow\phi$\;
  \For{$m \in M_{sys}$}{
    \textit{Step 1:} $m \leftarrow $  fit ($U$, $P$) \; 
    \textit{Step 2:} $P_{U-x} \leftarrow m(U-x)$\; 
    \textit{Step 3:} $\Delta P_x \leftarrow  P-P_{U-x}$\;
    $\varepsilon\leftarrow a\varepsilon \text{ of } m$\;
    $\rho\leftarrow corr(x, \Delta P_x)$\;
    \If{($M_{best}=\phi$) or ($\rho \geq \rho_{th}$ and $\varepsilon < \varepsilon_{Mbest}$) or ($\rho_{Mbest} < \rho_{th}$ and $\rho \geq \rho_{Mbest}$ and $\varepsilon_{Mbest} < \varepsilon_{Mbest}$)}{
        $M_{best}\leftarrow m$\;
    }
  }
  \If{$M_{best}\neq\phi$}{
    $\Delta P \leftarrow P - M_{best}(U-x)$
}
  \caption{Power model labeling}
  \label{algo:select}
\end{algorithm}

\subsection{Cross validation}
\label{subsec:model_validate}
When considering $k$ different dataset, let $\Delta P^{(i)}$ be the target container power label on testing dataset $i$ and $M^{(j)}(x^{(i)})$ be the predicted target container powerof dataset $i$ to the trained model from dataset $j$. The cross validation error is then calculated as 

\begin{equation}
\begin{split}
    \text{cross validation error } (c\varepsilon) &= \frac{\sum_{i=1}^k\sum_{j=1}^k\ c\varepsilon_{ij}}{k^2}\\
    c\varepsilon_{ij} =& error(\Delta P^{(i)}, M^{(j)}(x^{(i)})).
\end{split}
\end{equation}

\section{Evaluation Results}
\label{sec:experiment}
The experiments were conducted in three platform environments shown in Fig.~\ref{fig:base_sys}. The bare metal machine was an Intel x86-64 processor, equipped with an RAPL software power meter. Dynamic voltage and frequency scaling (DVFS) \cite{dvfs} was utilized to set two different CPU frequency maximum values and scaling governors. Minikube \cite{minikube} and Kubevirt \cite{kubevirt} were used to provide a Kubernetes container orchestration system, and a virtual machine as a container, respectively. 
We stressed the machine with three benchmark suites, each containing various kinds of workloads as listed in Table~\ref{tab:benchmark}, to validate the proposed container power model pertaining to the training biases \textit{P-I} and \textit{P-II} defined in Section~\ref{sec:problem}. 

Without losing generality, resource usage metrics were grouped by Kepler metric producers including (i) hardware counters, (ii) cGroups, (iii) BPF probe, and (iv) cAdvisor. The trainer module had six learning instances, namely, (i) linear regression, (ii) polynomial regression, (iii) k-nearest neighbors regression (kNN), (iv) gradient boosting regression (GBR), (v) stochastic gradient descent regression (SGD), and (vi) support vector regression (SVR). To mitigate complications in model training, we trained the model with the data set from the scenario of a single container running on bare metal. The data collected on virtual machine were used only for testing.

\begin{figure}[!t]
    \centering
    \includegraphics[width=0.4\textwidth]{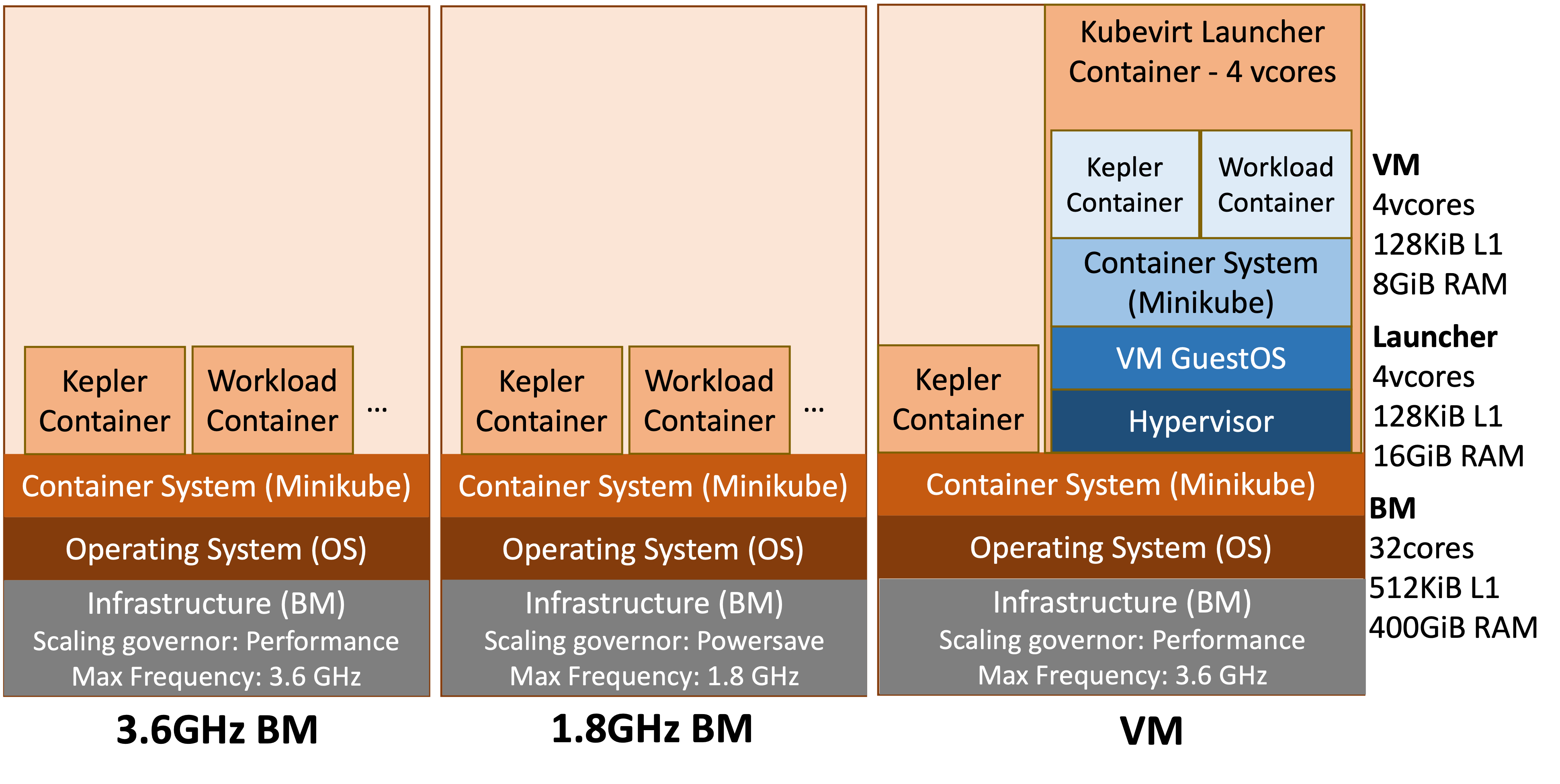}
    \caption{Experimental platform environment.}
    \label{fig:base_sys}
\end{figure}

 \begin{table}[!t]
    \caption{Benchmarks.}
    \label{tab:benchmark}
    \centering
    \begin{tabular}{l|l|c}
    \hline
    \textbf{Benchmark}&\textbf{Workload}&\textbf{Repetition}\\
    &&\textit{on BM/VM}\\
    \hline\hline
    Coremark \cite{coremark}&nthread x4,x8,x16,x32&10 / 3\\
    PARSEC \cite{parsec}&native bodytrack&5 / 3\\
    &native raytrace&\\
    &native canneal*&\\
    &native ferret&\\
    stress-ng \cite{stressng}&CPU x8,x16,x32 (30s)&10 / -\\
    &IO x8,x16,x32 (30s)&\\
    &Memory (2G) x8,x16,x32 (30s)& \\
    &CPU/IO/Memory (2G) x8,x16,x32 (30s)& \\
    \hline
    \multicolumn{3}{l}\footnotesize\textsuperscript{* native canneal cannot be tested on VM due to memory limitation.}\\
    \end{tabular}
\end{table}

\subsection{Comparison models}
\label{subsec:compare}
We compared our isolation method with the existing versions of non-RAPL power isolation illustrated in Fig.\ref{fig:proposed_concept}. Given $U$ as aggregated resource usage and $x$ as resource usage of the target container workload, the comparison models are explained as follows. 

\subsubsection{Proposed model $M$} $M$ uses the methodology described in Section \ref{sec:model}. Hence, 
\begin{equation}
\begin{split}
\textbf{training phase: }& M_{sys} = \text{fit}(U, P)\\
& M = \text{fit}(x, P-M_{sys}(U-x)),\\
\textbf{testing phase: }& M(x).\\
\end{split}
\end{equation}

\subsubsection{Profiling model \textbf{$M_{pf}$}} $M_{pf}$ uses the profiled background power ($P_{profile}$) to isolate the container power from the measured power. Hence, 
\begin{equation}
\begin{split}
\textbf{training phase: }& M_{pf} = \text{fit}(x, P-P_{profile}),\\
\textbf{testing phase: }& M_{pf}(x).\\
\end{split}
\end{equation}

\subsubsection{Heuristic model \textbf{$M_{dyn}$}} $M_{dyn}$ assumes the idle power equal to the power at the minimum point ($min(P)$). $min(P)$ is used for power isolation. The aggregated resource usage from all processes are used for training. Hence,
\begin{equation}
\begin{split}
\textbf{training phase: }& M_{dyn} = \text{fit}(U, P-min(P)),\\
\textbf{testing phase: }& M_{dyn}(x).\\
\end{split}
\end{equation}

\subsubsection{Model without isolation} Without isolation, zero idle power is assumed. The usage metrics from all processes and the measured power are used for training. This is equivalent to a system power modeling of $M_{sys}$. Hence,
\begin{equation}
\begin{split}
\textbf{training phase: }& M_{sys} = \text{fit}(U, P),\\
\textbf{testing phase: }& M_{sys}(x).\\
\end{split}
\end{equation}

To reduce bias in absolute number across different dataset, we normalized an error ($\varepsilon$) into percentage over the $\Delta P$ power range of each dataset, denoted $\%err$. 
$$
\%err = \frac{\varepsilon}{max(P)-P_{profile}}\times100
$$

\begin{figure}[!t]
\centering
\begin{subfigure}[b]{0.24\textwidth}
\centering
    \includegraphics[width=\linewidth]{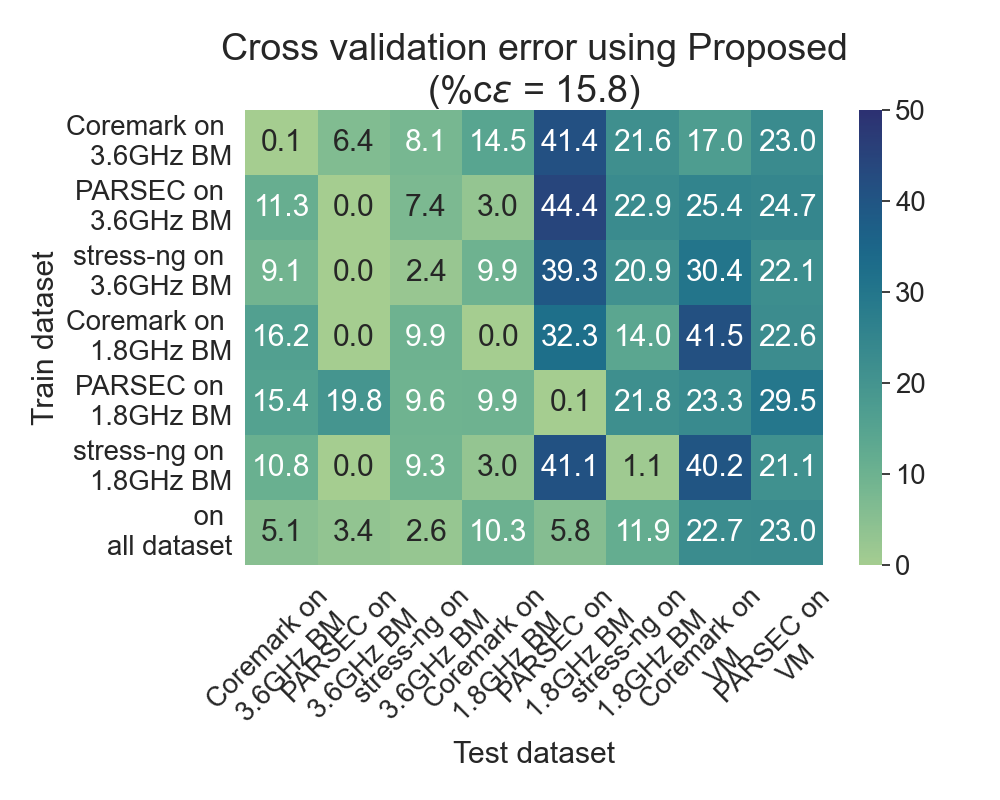}
    \subcaption{Proposed model}
    \label{fig:cross_proposed}
\end{subfigure}
\hfill
\begin{subfigure}[b]{0.24\textwidth}
\centering
    \includegraphics[width=\linewidth]{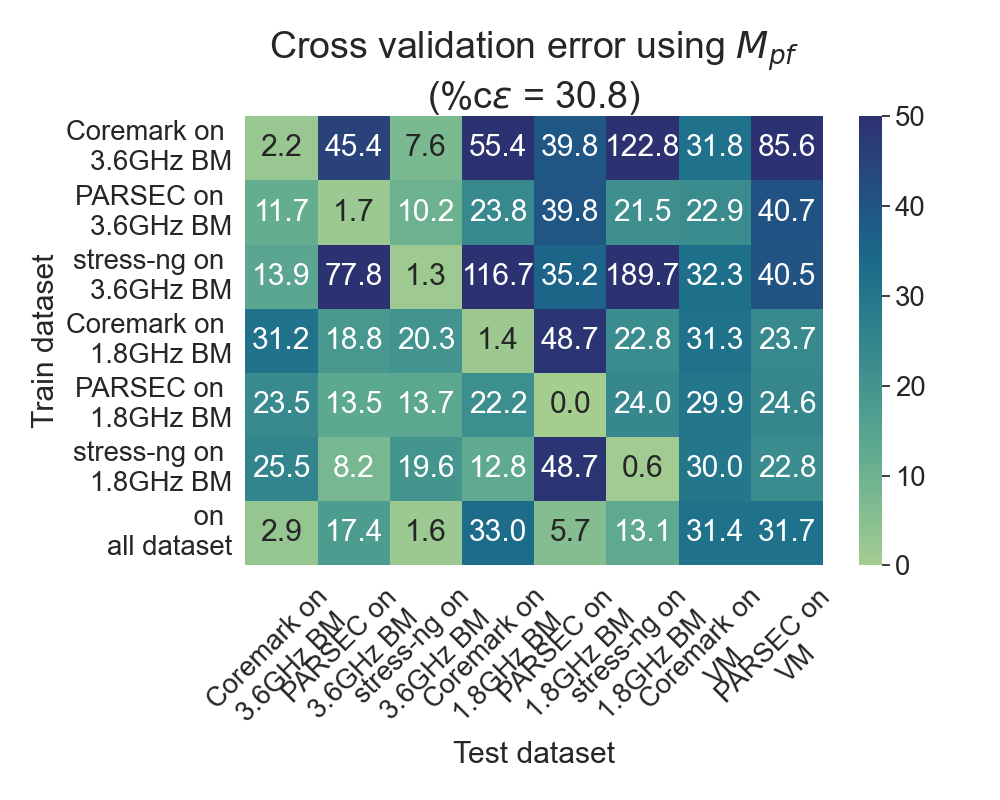}
    \subcaption{Profiling model}
    \label{fig:cross_profile}
\end{subfigure}
\hfill
\begin{subfigure}[b]{0.24\textwidth}
\centering
    \includegraphics[width=\linewidth]{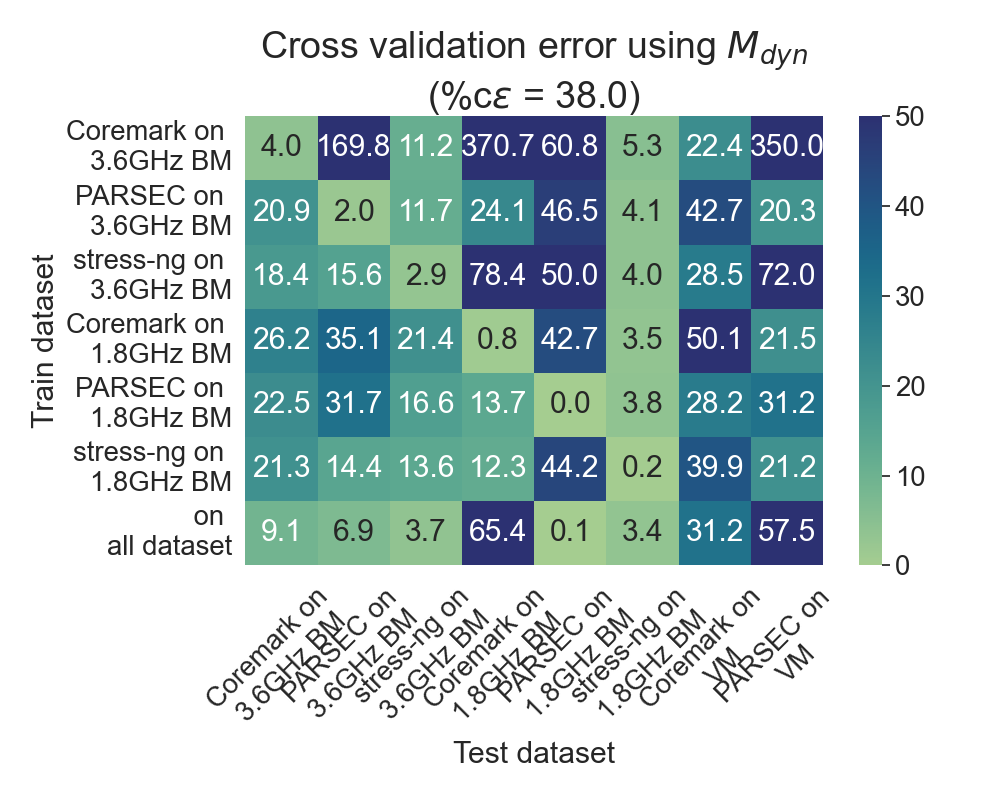}
    \subcaption{Heuristic model}
    \label{fig:cross_min}
\end{subfigure}
\hfill
\begin{subfigure}[b]{0.24\textwidth}
\centering
    \includegraphics[width=\linewidth]{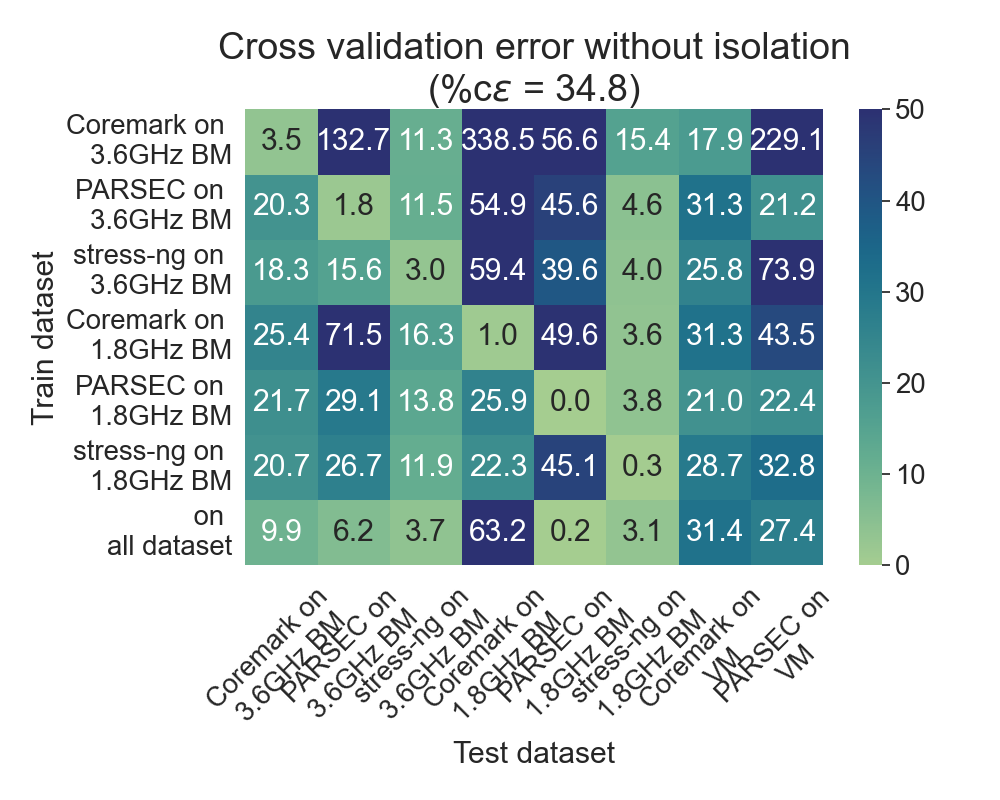}
    \subcaption{Power model without isolation}
    \label{fig:cross_none}
\end{subfigure}
\caption{Cross-validation error ($c\varepsilon$).}
\label{fig:cross_val}
\end{figure}

\subsection{Cross-workload, cross-platform validation}
\label{subsec:eval_cross}
The results of cross-workload, cross-platform validation are presented as heat map charts in Fig.~\ref{fig:cross_val}. A block ($i$, $j$) presents a cross validation error ($c\epsilon$) of the model trained by dataset $i$ on predicting the container power of dataset $j$. The last row presents prediction results using data from all datasets. 

The results leaded to the conclusion that the more we isolate a platform-specific power from the training process, the more accurate we can estimate a container power across different platforms and workloads ($c\varepsilon_{M} < c\varepsilon_{M_{pf}} < c\varepsilon_{M_{sys}}$). Particularly, the proposed model reduced an average cross validation error ($c\varepsilon$) in a half compared to the other models. For the heuristic model, the removed idle power was calculated from the power consumption when running benchmark workloads. Although it generally outperformed the model without isolation in diagonal blocks (i.e., same platform setting and workload), it was worst at predicting the container power across the workload and across the platform. With incremental training, all models had a lower cross-validation error as observed in the last row comparing to the other rows.

The above findings support the applicability of the proposed model to unseen containers on unknown platforms corresponding to problem definitions \textit{P-I} and \textit{P-II}. 

 \begin{table}[!t]
    \caption{Minimum power difference ($\Delta P_{min}$) and dynamic background power ($\Delta P_{bg}$).}
    \label{tab:sys_dep_power}
    \centering
    \begin{tabular}{l|c|c|c|c|c}
    \hline
         \textbf{Environment}&$P_{0}$&$P_{profile}$&\textbf{Benchmark}&$\Delta P_{min}$&$\Delta P_{bg}$\\
         \hline\hline
         3.6GHz BM&40.6&42.4&Coremark&23.9&107.2\\
         &&&PARSEC&43.9&61.8\\
         &&&stress-ng&10.7&81.9\\
         1.8GHz BM&12.8&26.5&Coremark&61.8&51.5\\
         &&&PARSEC&67.6&54.5\\
         &&&stress-ng&63.5&57.3\\
         VM&50.0&54.5&Coremark&28.6&118.8\\
         &&&PARSEC&41.6&46.9\\
         \hline
    \end{tabular}
\end{table}

In addition, we calculated a minimum power difference ($\Delta P_{min}$) and a dynamic background power ($\Delta P_{bg}$) for each dataset as shown in Table~\ref{tab:sys_dep_power}. $\Delta P_{min}$ is the incremental difference of the heuristic idle power, which was determined when the system was stressed by the benchmark, and the minimum power when running no process (i.e., profiled idle power, denoted by $P_0$). $\Delta P_{bg}$ is the increment of an average value of the predicted background power from an average value of the profiled background power, denoted by $P_{profile}$. $\Delta P_{min}$ was more than five times to the expected value, $P_0$, on the \textit{1.8GHz BM} platform environment. Meanwhile, $\Delta P_{bg}$ was more than two times to the profiled background power when running the Coremark benchmark. 

The findings in Table~\ref{tab:sys_dep_power} support the co-locating container concern of problem definition \textit{P-III} and resolve the dynamic background power identification of problem definition \textit{P-IV}.

\subsection{Isolation goodness ($\rho$)}
\label{subsec:eval_corr}
To clarify the importance of isolation goodness, we depict the prediction results from good and bad candidates of the system power models in Fig.~\ref{fig:isolate_detail}. Both results utilized the same features and learning approach (namely, cGroups metrics and gradient boosting regression). The only difference was the system power model features used in Fig.~\ref{fig:isolate_detail}(\subref{fig:bad_isolate}), which included the average CPU frequency. Although the candidate model in Fig.~\ref{fig:isolate_detail}(\subref{fig:bad_isolate}) achieved higher accuracy than the model in Fig.~\ref{fig:isolate_detail}(\subref{fig:good_isolate}), it had less correlated to the usage features. As a result, it was not able to isolate the background powers from the measured power. In contrast, the correlation between power after isolation and target workload usage can still be observed in the third sub-graph in Fig.~\ref{fig:isolate_detail}(\subref{fig:good_isolate}). 

Correspondingly, we also observed that 40\% of the isolated data from the proposed model and those from the profiling model had a high isolation goodness which is more than 0.7. Whereas, the heuristic model can produce only 27\% of isolated data that had a high isolation goodness.

\begin{figure}[t!]
\centering
\begin{subfigure}[b]{0.24\textwidth}
    \includegraphics[width=1\linewidth]{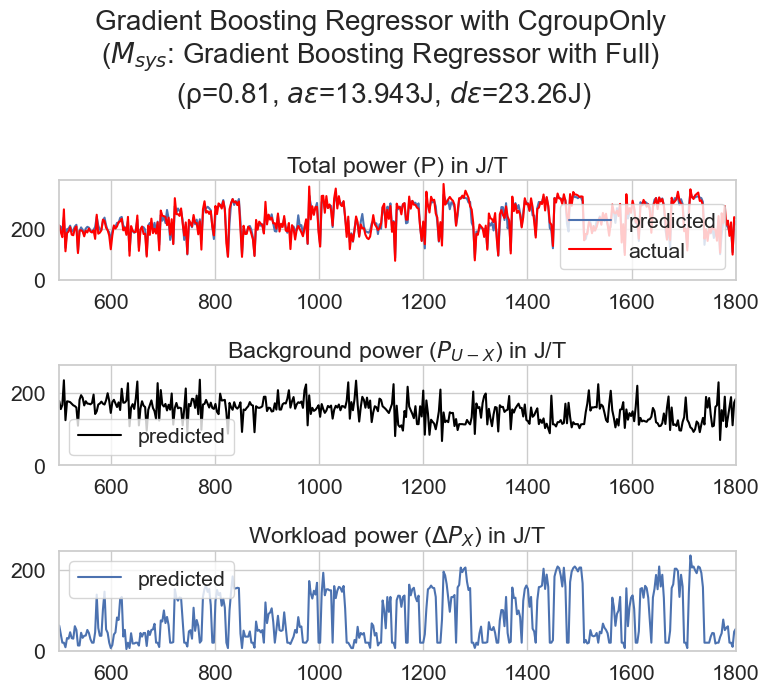}
    \subcaption{Candidate with good isolation (high correlation)}
    \label{fig:good_isolate}
\end{subfigure}
\hfill
\centering
\begin{subfigure}[b]{0.24\textwidth}
    \includegraphics[width=1\linewidth]{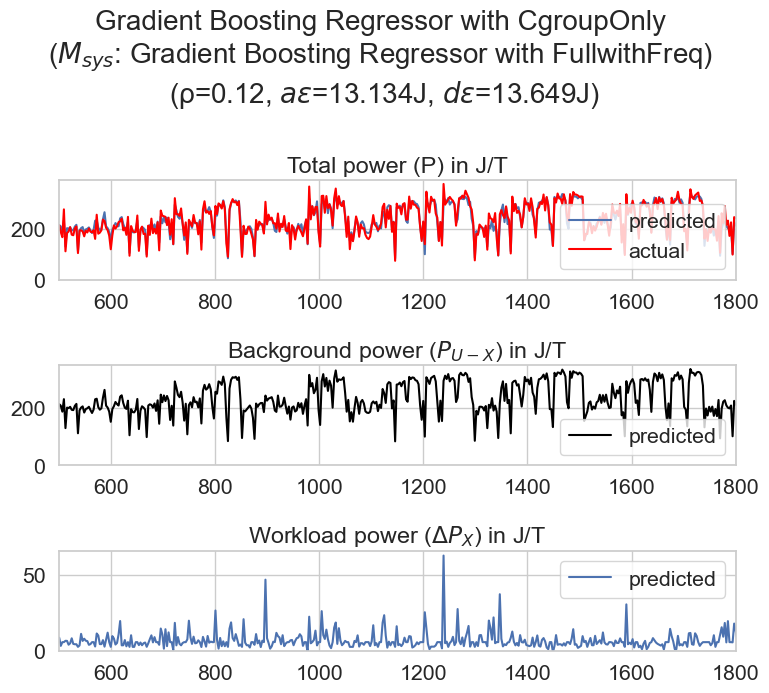}
    \subcaption{Candidate with bad isolation (low correlation)}
    \label{fig:bad_isolate}
\end{subfigure}
\caption{Sample results of isolation goodness ($\rho$) vs. prediction error (a$\varepsilon$) from Coremark.}
\label{fig:isolate_detail}
\end{figure}

In addition, the selected system power model was also applicable in the noisy scenario, as shown in Fig.~\ref{fig:noisy_experiment}(\subref{fig:noisy_usage}). The noisy background power during t = 200s and t = 800s can be uncovered and removed from the container power model training, as presented in Fig.~\ref{fig:noisy_experiment}(\subref{fig:noisy_predict}). 

\begin{figure}[t!]
\centering
\begin{subfigure}[b]{0.24\textwidth}
    \includegraphics[width=\linewidth]{fig/noisy_workload.png}
    \subcaption{Resource usage of Coremark}
    \label{fig:noisy_usage}
\end{subfigure}
\hfill
\centering
\begin{subfigure}[b]{0.24\textwidth}
    \includegraphics[width=\linewidth]{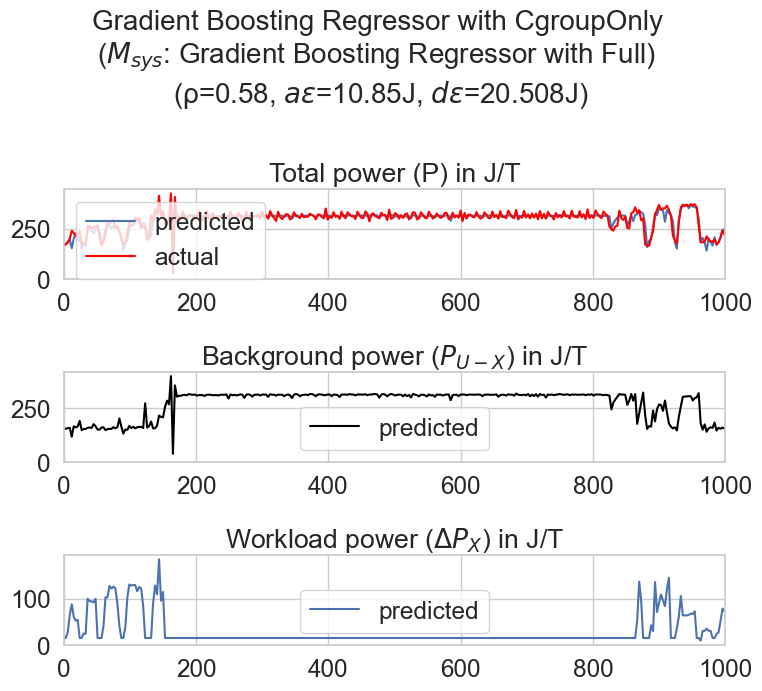}
    \subcaption{Power isolation result}
    \label{fig:noisy_predict}
\end{subfigure}
    \caption{Proposed power isolation in noisy scenario.}
    \label{fig:noisy_experiment}
\end{figure}

\section{Conclusion}
\label{sec:conclude}
This paper introduced a pipeline framework for training a container power model. The proposed framework allows Kepler, a Cloud-native energy-related metric exporter, to estimate individual container power consumption in the unknown platforms that have no access to an online power measurement. We highlighted the isolator module in the pipeline, which estimates a power consumed by workload containers used as model training labels. We proposed a new isolation approach considering a dynamic background power, which cannot be profiled in advance. In addition, we also defined a new metric to determine a goodness of isolation. 
The evaluation results showed that the proposed method can improve a cross-workload, cross-platform prediction accuracy two times to the comparable isolation methods.

\bibliographystyle{IEEEtran}
\bibliography{references}

\end{document}